\documentclass[runningheads]{svmult}

\usepackage{makeidx}   
\usepackage{graphicx}  
\usepackage{physprbb}  
\usepackage[square,comma,numbers,sort&compress,sectionbib]{natbib}
\usepackage{amsmath}   
\usepackage{amsfonts}  %
\usepackage{amssymb}   %
\usepackage{latexsym}  
\newcommand{\ket}[1]{\mbox{$|#1\rangle$}}
\newcommand{\bra}[1]{\mbox{$\langle#1|$}}

\begin{document}
\title*{Decoherence in Discrete Quantum Walks}
\toctitle{Decoherence in Discrete Quantum Walks}
\titlerunning{Decoherence in Discrete Quantum Walks}
\author{Viv Kendon \and Ben Tregenna}
\authorrunning{Viv Kendon \and Ben Tregenna}
\institute{QOLS, Imperial College London, Blackett Laboratory, London, SW7 2BW, UK}

\maketitle              

\begin{abstract}
We present an introduction to coined quantum walks on
regular graphs, which have been developed in the past few years
as an alternative to quantum Fourier transforms for underpinning 
algorithms for quantum computation.
We then describe our results on the
effects of decoherence on these quantum walks 
on a line, cycle and hypercube.
We find high sensitivity to decoherence, increasing with the number
of steps in the walk, as the particle is becoming more delocalised
with each step.
However, the effect of a small amount of decoherence 
can be to enhance the properties of the quantum walk that are desirable for
the development of quantum algorithms, such as fast mixing times to
uniform distributions.
\end{abstract}

\section{Introduction to Quantum Walks}\label{qrwsec}

Quantum walks are based on a generalisation of 
classical random walks, which have found many 
applications in the field of computing.
Examples of the power of classical random walks to solve hard
problems include algorithms for solving $k$-SAT
\cite{schoning99a}, estimating the volume of a convex body \cite{dyer91a}, and
approximation of the permanent \cite{jerrum01a}.
They are a subset of a wider model of computation, cellular automata,
which have been proved universal for classical computation.
The utility of classical walks suggests that 
extending the formalism to the quantum regime may assist the new
field of quantum information processing in generating further quantum
algorithms. Similarly to the classical case, it is also possible to
define the notion of quantum cellular
automata, whose equivalence to quantum Turing machines has been shown
\cite{watrous95a}.

Most known quantum algorithms are based on the {\em quantum
Fourier transform}, for an introduction to quantum computing and algorithms
see, e.~g., \cite{shor00a,nielsenchuang00}.
Quantum versions of random walks provide a distinctly different paradigm
in which to develop quantum algorithms.  Very recently, two
such algorithms have been presented.  Shenvi {\em et al.}~\cite{shenvi02a}
proved that a quantum walk can perform the same task as Grover's search
algorithm \cite{grover96a}, with the same quadratic speed up.
Childs {\em et al.}~\cite{childs02a} describe a quantum algorithm
for transversing a particular graph exponentially faster than
can be done classically.  This exponential speed up is very
promising, though the problem presented is somewhat contrived.

In fact, several possible
extensions of classical random walks to the quantum
regime have been proposed \cite{farhi98a,childs01a,grossing88a},
however, here we will only treat the \textit{discrete time, coined
quantum walks} \cite{aharonov00a}, subsequently these are referred to simply
as ``quantum walks''.  Before introducing quantum walks, it is
helpful to review the properties of classical random walks.
This is followed by an overview of
quantum walks on a line, $N$-cycle, and hypercube.  Section \ref{qrwdec}
presents our results on the effects of decoherence in these quantum walks.

\subsection{Classical Random Walks on Graphs}

The discrete space on which a random walk takes place can most generally
be described as a graph $G(V,E)$ with two components, a set of
vertices $V$, and a set of edges $E$.  An edge may be specified by the
pair of vertices that it connects, $e=(v_i,v_o)$.  The graph is {\em undirected}
when $(v_i,v_o)\in E ~{\text{iff}} (v_o,v_i)\in E$. The second
essential feature of a classical walk is the (time-independent)
transition matrix $M$, whose elements $M_{ij}$ provide the probability
for a transition from vertex $v_i$ to vertex $v_j$.
These probabilities are non-zero
only for a pair of vertices connected by an edge,
\begin{equation}
M_{ij}\neq 0 \mbox{\hspace{1em} iff \hspace{1ex}} e=(v_i,v_j)\in E\;.
\end{equation}
The walk is {\em unbiased} if the non-zero
elements of $M$ are given by $M_{ij}=\frac{1}{d_i}$, where $d_i$ is
the degree of the vertex $v_i$ (the number of edges connected to
$v_i$). A graph is called {\em regular ($d$-regular)}
if all vertices have equal degree ($d$).
The state of a classical random walk at a given time $t$ is
described by a probability distribution $P(v,t)$ over the vertices
$v\in V$.  This distribution evolves at each time step by application of the
transition matrix $M$,
\begin{equation}
P(v,t)=M^t P(v,0)\;.
\end{equation}

A number of features of these classical walks are worthy of note for
later comparisons. If $G$ is {\em connected} (every pair of vertices have a path
linking them via a sequence of edges), then the walk tends to a steady state
distribution $\pi$ which is independent of the initial state
$P(v,0)$.  Further, if the graph is regular then the limiting distribution
is uniform on all the vertices.  An exception arises in the case of
periodic random walks, however, a ``resting probability''
for the walk to remain at the current vertex may be added which breaks
the periodicity and restores the usual convergence. The rate of
convergence to this limiting distribution may described in a number of
ways, here we choose the \textit{mixing time},
\begin{equation}\label{cme}
M_{\varepsilon}^{\mathrm{C}}=\min\left\{T|\forall t>T: ||P(v,t)-\pi||_{\text{tv}}<\varepsilon\right\}\;.
\end{equation}
The measure used here is the \textit{total variational distance},
\begin{equation}
||p_1-p_2||_{\text{tv}}=\sum_{v_i\in V} |p_1(v_i)-p_2(v_i)|\;.
\end{equation}
The mixing time thus gives a measure of the time after which the
distribution is within a distance $\varepsilon$ of the limiting distribution
and remains at least this close.
An alternative measure that is useful in classical algorithms, for
example 3-SAT \cite{schoning99a}, is the \textit{hitting time}. This is defined
for a pair of vertices $v_0$, $v_1$ as the expected time at which a walk
starting at $v_0$ reaches $v_1$ for the first time.

\subsection{Criteria for a Quantum Walk}

Given a $d$-regular graph, $G(V,E)$, the associated Hilbert space may
be defined as
\begin{equation}\label{ptclHsp}
{\cal H}_V ={\text{span}} \{\ket{v_i}\}^{|V|}_{i=1}\;.
\end{equation}
As already mentioned, several extensions of classical random walks
to the quantum regime have been proposed, there is no unique way to do this.  
To guide us, we note there are several properties of classical random walks and
quantum systems that we would like such quantum walks to have, if
possible.  The three desirable properties of the classical transition matrix
$M$ are,
\begin{itemize}
\item locality: $M_{ij}\neq 0 ~{\text{iff}} ~e=(v_i,v_j)\in E$, i.~e.,
	transitions are only between vertices directly connected by an edge
\item homogeneity: if $e=(v_i,v_j)\in E, ~ |M_{ij}|=\frac{1}{d_i}$, i.~e.,
	equal probability of transition to any neighbouring vertex 
\item time independence: the current step does not depend
	in any way on previous steps of the random walk (Markovian)
\end{itemize}
The quantum transition matrix $U$, equivalent to $M$, should have
similar properties to $M$.  In addition, it should be
\begin{itemize}
\item unitary
\end{itemize}
because all pure quantum processes are unitary.
As proved by Meyer \cite{meyer96a,meyer96b}, this requirement of unitarity is
incompatible with the three prior properties, except in very special
circumstances that don't produce interesting quantum dynamics.
In order to generate a non-trivial quantum evolution, one or more
of these constraints must be relaxed.
The continuous time quantum walks proposed by Farhi and Gutmann
\cite{farhi98a}, is non-local, in the sense that there is a small
probability of the particle moving arbitrarily far away in a given unit
time interval.  The quantum cellular automata of Meyer \cite{meyer96b,meyer96c}
are not homogeneous, in that the full dynamics is specified over
two time steps rather than one.
One may make the quantum walk non-unitary by measuring the particle at each
step, but this simply reproduces the classical random walk.
Finally, one may consider relaxing the time independence condition
slightly, and this is what the coined quantum walks effectively do.
For completeness, we note that there is an equivalent formulation
of the coined quantum walk in terms of a simple quantum process on a
directed graph, derived from the original undirected graph, that is due to
Watrous \cite{watrous01a}.

\subsection{Coined Quantum Walks}

By analogy with the classical walk, in which one
pictures flipping a coin at each time-step to determine which edge to
leave the current vertex by, interesting quantum results may be
obtained by the introduction of an explicit (quantum) coin.
In addition to the particle, with associated Hilbert space
${\cal H}_{\mathrm{V}}$ as given in (\ref{ptclHsp}), there is
an auxiliary quantum system, the coin, with a Hilbert space of dimension $d$
(the degree of the graph) ${\cal H}_{\mathrm{C}}=\mathbb{C}^d$.
The transition matrix is then defined as a unitary matrix $U$
acting on the tensor product of these two spaces.  This unitary is
constructed from two separate operators, a ``coin flip'' and a
conditional translation. For an unbiased walk, the coin flip operator,
$\mathbf{C}$ is a unitary matrix whose elements all have equal modulus in the
computational basis of the coin. This adds significant new degrees
of freedom to the system, as the relative phases of these elements may
be chosen arbitrarily. 
The conditional translation $\mathbf{T}$ moves the particle along an edge to
an adjacent vertex determined by the state of the coin.
The evolution of the system from an initial state $\ket{\psi(0)}_{\mathrm{CV}}$
to $\ket{\psi(t)}_{\mathrm{CV}}$ after $t$ steps is thus given by,
\begin{equation}
\ket{\psi(t)}_{\mathrm{CV}}=\left[\mathbf{T}\cdot(\mathbf{C}\otimes
\mathbb{I})\right]^t\ket{\psi(0)}_{\mathrm{CV}}\;.
\label{coinevol}
\end{equation}
The coin can also be thought of as forming a $d$-state quantum memory
from one step to the next, allowing a much wider range of dynamics to
be fully reversible (a necessary property of unitary evolution).

Unitary evolution is also completely deterministic, so the choice of initial
condition is never ``washed out'', rather, it plays a key role in
determining the outcome of the quantum walk.
Unitarity also means that the joint state of the coin and
particle can never reach a steady state.  Even the induced probability
distribution over the nodes obtained by tracing out the coin,
\begin{equation}
P(v,t)=~_{\mathrm{V}}\bra{v}
{\mathrm{Tr_C}}\left[\ket{\psi(t)}_{\mathrm{CV}}\bra{\psi(t)}\right]\ket{v}_{\mathrm{V}}\;,
\label{Pdef}
\end{equation}
does not converge to a long-time limit.
However, it is possible to define a time-averaged probability
distribution that does reach a steady state,
\begin{equation}
\overline{P(v,T)}=\sum_{t=0}^{T-1} \frac{P(v,t)}{T}\;.
\end{equation}
Operationally, this is easy to produce, it is the distribution obtained by
sampling the particle location at some time $t$ uniformly selected at
random from $0\le t<T$.
Using this distribution it is possible to define a quantum mixing time
similar to that in (\ref{cme}),
\begin{equation}\label{qme}
M_{\varepsilon}^{\mathrm{Q}}=\min\left\{T|\forall t>T: ||\overline{P(v,t)}-\pi||_{\text{tv}}<\varepsilon\right\}\;.
\end{equation}

The second measure discussed for classical random walks was the
hitting time. This can also be extended to quantum walks, but as
before will require some modification. In the classical case it is
possible to measure the location of the particle at each time step to
ascertain whether or not it has reached the desired vertex. The
equivalent quantum measurement disturbs the system; if a complete
projection onto the vertex Hilbert space is performed at each step
then all coherences are lost and a classical distribution
results. There are two approaches to this problem that have been used
in the literature \cite{kempe02a}. It would be possible to wait until
a chosen time, $T$, and then perform a full measurement over all the
vertices. If the probability of being at the desired vertex is greater
than some value $p$ which is bounded below by an inverse polynomial in
the size of the graph, it is said that the walk has a $(T,p)$
one-shot hitting time.
(This lowerr bound ensures that standard amplification procedures can
efficiently raise the success probability arbitrarily close to unity.)
The second alternative is to perform a partial
measurement at each time step. Projecting onto the subspaces given by
the desired vertex $v$, $I\!\!P_v=\ket{v}\bra{v}$, and its orthogonal
complement, $I\!\!P_{\bar{v}}=\mathbb{I}-\ket{v}\bra{v}$ will halt the
walk once the vertex $v$ is reached. A walk has a $(T,p)$ concurrent
hitting time if $I\!\!P_v$ is measured with probability greater or
equal to $p$, which again must have as a lower bound an inverse
polynomial in the graph size, in a time $T$. Both these quantities
have been shown to exhibit an exponential speed up over the classical
case for a walk between opposite corners of a hypercube
\cite{kempe02a}. 

Explicit solution of a quantum walk has proved to be a difficult
problem. In certain cases standard techniques from classical graph
theory have been applied with some success, namely solution in the
Fourier space of the problem \cite{ambainis01a,aharonov00a}, and also the technique of
generating functions \cite{bach02a}. Fourier solutions are possible when
the graph is of a particular form, known as a Cayley graph. Any
discrete group has an associated Cayley graph, in which the elements
of the group form the vertices and the edges are placed by choosing a
complete set of generators, $g_i$ and placing an edge between the
vertices $a$ and $b$ iff, $a=g_i b$ for one of the $g_i$.  This produces
an undirected graph results if the group is Abelian.
We will next consider three specific cases of such graphs for which analytical
solution has proved possible.

\subsection{Coined Quantum Walk on a Line}

\begin{figure}[t!]
    \begin{center}
	\includegraphics[width=0.7\columnwidth]{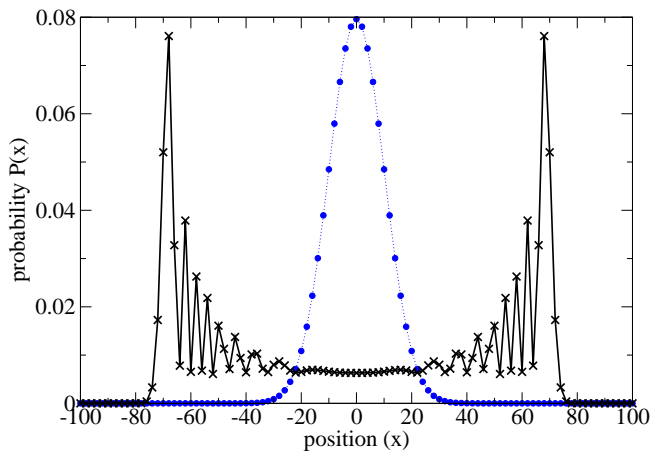}
    \end{center}
    \caption{A comparison of the probability distribution of a classical
	and coined quantum walk on a line, after 100 time steps. Only even
	points are plotted as both functions are zero on the odd grid points}
    \label{qrw100pdf}
\end{figure}

Consider the case when the graph consists of an lattice of points at
integer positions on an infinite line. The Hilbert space of the particle is
represented by the integers,
\begin{equation}
{\cal H}_V={\text{span}} \{\ket{x}:x\in \mathbb{Z}\}\;.
\end{equation}
At each time interval the particle can move one step to either the
left or the right, the coin thus requires only a two dimensional
Hilbert space which can conveniently be written as,
\begin{equation}
{\cal H}_C=\{\ket{R},\ket{L}\}\;.
\end{equation}
The conditional translation $\mathbf{T}$ thus has the following action on
the basis states,
\begin{equation}
\mathbf{T}\ket{R,x}=\ket{R,x+1}\;,\hspace{0.75cm}
\mathbf{T}\ket{L,x}=\ket{L,x-1}\;. 
\end{equation}
The evolution of the probability distribution induced on the lattice
may now be compared with that of the analogous classical walk on a line.
It is sufficient to us the Hadamard operator as the coin flip,
\begin{equation}
\mathbf{H}=\frac{1}{\sqrt{2}}\left( \begin{array}{cc}
        1&1\\
        1&-1
        \end{array} \right)\;,
\label{haddef}
\end{equation}
all others are essentially equivalent \cite{ambainis01a,bach02a}.
We also choose the initial state to be
\begin{equation}
\ket{\psi(0)}=\frac{1}{\sqrt{2}}\left(\ket{R}+i\ket{L}\right)\otimes\ket{0}\;.
\end{equation}
This choice produces a symmetric distribution, as in the classical case.
Since $\mathbf{H}$ and $\mathbf{T}$ contain only real elements, 
there is no interference between the part of the walk due to the
initial state \ket{R} and that due to $i\ket{L}$.  Thus the natural
(in terms of basis states) outcome of a quantum walk is actually biased,
unlike a classical random walk, a basic demonstration of the initial
condition affecting the entire outcome of the walk.
\begin{figure}[t!]
    \begin{center}
	\includegraphics[width=0.7\columnwidth]{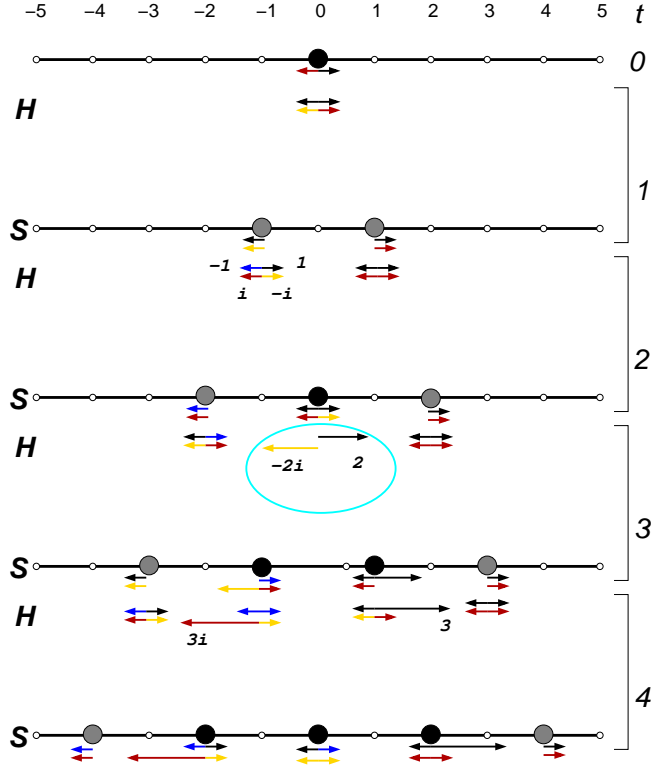}
    \end{center}
    \caption{A schematic representation of the first four steps of the
	evolution of the 1-D walk with a Hada\-mard coin.
	Circles show the particle positions and arrows show the state
	of the coin, with relative magnitudes indicating the amplitude
	of the components of the wavefunction}
    \label{qrwline}
\end{figure}
Four steps of this evolution are shown 
schematically in Fig.~(\ref{qrwline}) and the respective probability
distributions are shown in Fig.~(\ref{qrw100pdf}) after 100 time steps.
The classical walk forms a binomial distribution, with a mean
of 0 and standard deviation $\sigma_{\mathrm{C}}=\sqrt{T}$. Figure
\ref{qrw100pdf} displays several of the interesting features of a quantum walk. 
The central interval $[-2\sqrt{T}, 2\sqrt{T}]$ is essentially uniform in the
quantum case, with oscillating peaks outside this region up to
$[-T/\sqrt{2}, T/\sqrt{2}]$.
This is due to interference between the large number of possible paths to each
point in this range.

The quantum walk on a line has been solved exactly \cite{ambainis01a,nayak00a}
using both real space (path counting) and Fourier space methods.
The solutions are complicated, mainly due to the
``parity'' property, namely, that the solutions must have support
only on even(odd)-numbered lattice sites at even(odd) times.
The moments can be calculated for asymptotically large times, for a
walk starting at the origin, the standard deviation is
\begin{equation}
\lim_{T\rightarrow\infty} \sigma_{\mathrm{Q}}=\left( 1-\frac{1}{\sqrt{2}}\right)^{1/2} T\;.
\end{equation}
The standard deviation $\sigma_{\mathrm{Q}}$
is thus linear in $T$, in contrast to $\sqrt{T}$ for the classical walk.

Quantum walks on a line have now been studied in considerable detail.
Discussions of absorbing boundaries have been given 
\cite{bach02a,yamasaki02a,ambainis01a}, with applications to halting
problems in mind.  Extensions to multiple coins have been made by
Brun \textit{et al.}~\cite{brun02a,brun02c}.  However, though useful for
understanding the basic properties of quantum walks, the walk
on a line is too simple to yield interesting quantum problems for significant
algorithms.

\subsection{Coined Quantum Walk on a $N$-Cycle}\label{ex_lorc}

If periodic boundary conditions are applied, instead of a walk on an
infinite line, a closed walk on a $N$-cycle is obtained. For this system,
which wraps round on itself, the standard deviation is inappropriate and
mixing times must be considered.  The classical result is well known,
\begin{equation}
\overline{M}_{\varepsilon}^{\mathrm{C}}=o(N^2 /\varepsilon)\;.
\end{equation}
Here the time average of the probability distribution, $\overline{P(v,T)}$
has been used in (\ref{cme}) for ease of comparison with the quantum case.
The usual mixing time defined by (\ref{cme}) using $P(v,T)$ scales as
$M_{\varepsilon}^{\mathrm{C}}=o(N^2\log(1/\varepsilon))$.
The scaling of $\varepsilon$ is not important for the quantum--classical
comparisons we will do here, more details can be found in \cite{aharonov00a}.
The equivalent quantity for a quantum walk was bounded above by
Aharonov \textit{et al.}~\cite{aharonov00a}, and shows a quadratic speed-up,
\begin{equation}
M_{\varepsilon}^{\mathrm{Q}}\leq O\left(\frac{N \log N}{\varepsilon^3}\right)\;.
\end{equation}
Bounds have also been established in the case of more general graphs
and it is conjectured that mixing times can be improved at most
polynomially by quantum walks \cite{aharonov00a}.
Numerical studies, see \cite{kendon02b}, suggest
the form of the quantum mixing time is actually
$M_{\varepsilon}^{\mathrm{Q}}= O\left(N/\varepsilon\right)$,
and explain why tighter analytical bounds are hard to obtain.

\subsection{Coined Quantum Walk on a Hypercube}\label{ex_hype}

A hypercube of dimension $N$ is a Cayley graph with $2^N$
vertices labelled by the bit-strings of length $N$.
The Hilbert space for this graph is spanned by the basis states
$\ket{x},\, x\in[0,2^N]$.
Each vertex has degree $N$, so the state space of the coin is
$\mathcal{H}_{\mathrm{C}}=\mathbb{C}^N$.
The basis vectors of this space are denoted by
$\ket{a},\, a\in[0,N]$.  These states correspond to the $N$ vectors
$\ket{e_a}$ where $e_a$ is the vector with all zeroes except for a
single one in the $a^{\text{th}}$ position.
The translation operator for the walk on a hypercube can then be defined as,
\begin{equation}
\mathbf{T}\ket{a,x}=\ket{a,x\oplus e_a}\;.
\end{equation}
Any $N\times N$ unitary matrix with all elements of unit modulus may be used
as an unbiased coin, however, this is not the most natural choice
given the symmetries of the hypercube.  Instead, a biased coin has been
selected, which distinguishes the edge along which the particle arrived
at the vertex from all the others \cite{moore01a}.
This ``Grover'' coin acts thus,
\begin{equation}
\mathbf{G}\ket{a}=\frac{2}{N}\sum_b \ket{b} -\ket{a}\;.
\label{grovercoin}
\end{equation}
The full evolution for a single time step of this walk is then given
by $\mathbf{U}=\mathbf{T}\cdot(\mathbf{G}\otimes\mathbb{I})$.

The quantum walk on the hypercube can be solved analytically by
mapping it to a walk on a line with a variable coin operator,
provided the symmetry of the hypercube is maintained throughout.
To preserve the symmetry between particle states with equal Hamming
weights, defined to be the number of 1's in their bit-string, the
initial state must be chosen to be localised on the hypercube at
\ket{0} with the coin in an equal superposition of all its states,
$\ket{\psi(0)}=\tfrac{1}{\sqrt{N}}\sum_a \ket{a} \otimes \ket{0}$.
With this ansatz it has been shown by Moore and Russell \cite{moore01a}
that the walk has an exact \textit{instantaneous} mixing time,
when the distribution transiently becomes exactly uniform, at $T=\pi N/4$.
More importantly, Kempe \cite{kempe02a} found
the first exponential gap between classical and quantum walks 
for hitting times on the hypercube.  Both the
concurrent and one-shot hitting times for a walk to travel from
\ket{0} to \ket{2^N} are found to be $\pi N/2$, with small error
probability. This is exponentially faster than the classical hitting
time, $T=2^{N-1}$.  This is not a true quantum speed up, since there
are more sophisticated classical algorithms than a random walk to
reach the opposite corner that exploit the symmetry of the hypercube.
Nonetheless, it is important as the first indication that quantum walks
have the potential for driving quantum algorithms with exponentially
faster, later confirmed by Childs \textit{et al.}~\cite{childs02a}
on a more random graph.

\section{Decoherence in Quantum Walks}\label{qrwdec}

We will now study each of the three systems just introduced 
in the presence of decoherence. A simple model of decoherence is
chosen here, at each time step of the quantum walk a measurement is
made in the computational basis with a probability $p$. We consider
three cases, where the measurement is over only the coin degrees of
freedom, only over the particle states, or is a complete measurement of both.
Given a unitary transform $\mathbf{U}=\mathbf{T}\cdot(\mathbf{C}\otimes
\mathbb{I})$ for each step of the walk as
described by (\ref{coinevol}), the effect of this decoherence model
can be described as a discrete master equation,
\begin{equation}\label{decme}
\varrho(t+1) = (1-p)U\varrho(t) U^\dagger + p \sum_i I\!\!P_i U\varrho(t)
U^\dagger I\!\!P_i\;.
\end{equation}
The summation runs over the dimensions of the Hilbert space on which
the decoherence occurs, either the coin $\mathcal{H}_C$, the particle
$\mathcal{H}_V$ or both $\mathcal{H}_C \otimes \mathcal{H}_V$. The
projectors $I\!\!P_i$ are defined to act in the computational basis.
When $p=0$ the ideal quantum walk
is obtained, and for $p=1$ when a measurement is made at
each time step it produces the classical random walk.

The numerical work summarised in the remainder of this paper
has been presented more fully in \cite{kendon02b}.

\subsection{Decoherence in a Quantum Walk on a Line}

Since there is an experimental proposal to implement a quantum walk on a line
in an optical lattice \cite{dur02a}, as well as the three examples for
$I\!\!P_i$ given above, we considered the likely form of experimental errors,
and also modeled the effect of an imperfect Hadamard on the coin.
The Hadamard operation may be considered to be a ``rotation''
about the computational basis by $\pi/4$, (actually a rotation and
reflection since $\det(\mathbf{Rot}(\theta))=-1$),
\begin{equation}
\mathbf{Rot}(\theta)=\left( \begin{array}{cc}
        \sin(\theta)&\cos(\theta)\\
        \cos(\theta)&-\sin(\theta)
        \end{array} \right)\;,
\end{equation}
The error model used in this case consisted of a Gaussian spread of
standard deviation $\sqrt{p}\pi/4$ about the ideal value of
$\theta=\pi/4$.

\begin{figure}[t!]
    \begin{center}
	\includegraphics[width=0.7\columnwidth]{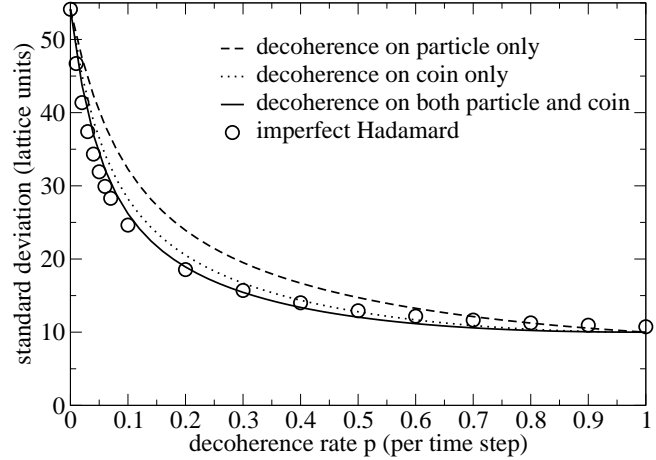}
    \end{center}
    \caption{Standard deviation $\sigma_p(T)$ of the particle on
	a line for different models of decoherence, for $T= 100$ time steps}
    \label{linedecall}
\end{figure}

All types of decoherence model produce the same general form for the decay of 
$\sigma_p(T)$ from the quantum to the classical value,
with small differences in the rates, as shown in Fig. \ref{linedecall}.
The slope of $\sigma_p(T)$ is finite as $p\rightarrow 0$ and zero
at $p = 1$.  We have calculated $\sigma_p(T)$ analytically
for $pT \ll 1$ and $T \gg 1$ for the case where $I\!\!P_i$ is the
projector onto the preferred basis $\{\ket{a,x}\}$, i.~e., the decoherence
affecting both particle and coin. Details are given in \cite{kendon02c}),
the result is
\begin{equation}
\sigma_p(T) \leq \sigma(T)\left[1-\frac{pT}{6\sqrt{2}} +
\frac{p}{\sqrt{2}}(1-1/\sqrt{2}) + O(p^2, 1/T)\right]\;.
\end{equation}
The first order dependence is thus proportional to $pT$, 
so the sensitivity to decoherence grows linearly
in $T$ for a given decoherence rate $p$. 

\begin{figure}[t!]
    \begin{center}
	\includegraphics[width=0.7\columnwidth]{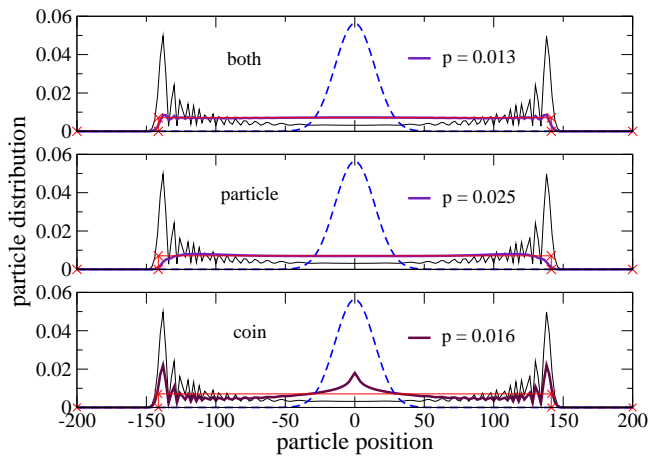}
    \end{center}
    \caption{Distribution of the particle position for a quantum walk on a
	line after $T=200$ time steps. Pure quantum (dotted), fully
	classical(dashed), and decoherence at rate shown on part of system
	indicated by key (solid). Uniform distribution between $-T/\sqrt{2}
	\le x \le T/\sqrt{2}$ (crosses) also shown} 
    \label{linedist}
\end{figure}

There are interesting differences in the shape of the distribution
of the particle position for each of the types of decoherence.  
The decoherence rate that gives the closest to uniform distribution has been
selected and plotted in Fig. \ref{linedist},
along with the pure quantum and classical distributions
for comparison.
When the particle position is subject
to decoherence that tends to localise the particle in the standard basis,
this produces a highly uniform distribution between $\pm T/\sqrt{2}$ for a
particular choice of $p$.  The optimal decoherence rate $p_u$ can be obtained
by minimising the total variational distance between the actual and
uniform distributions,
\begin{equation}
\nu(p,T) \equiv ||P(x,p,T) - P_u(T)||_{\text{tv}} \equiv \sum_x|P(x,p,T) - P_u(T
)|\;,
\end{equation}
where $P(x,p,T)$ is the probability of finding the particle at position $x$
after $T$ time steps, regardless of coin state [compare (\ref{Pdef})],
and $P_u(T) = \sqrt{2}/T$ for $-T/\sqrt{2} \le x \le T/\sqrt{2}$
and zero otherwise.
The optimum decoherence rate depends on the number of steps in the walk,
it can be determined numerically that $p_uT \simeq 2.6$ for decoherence on
both the particle and the coin,
and $p_uT \simeq 5$ for decoherence on the particle only.
These differences in the quality of the uniform distribution are 
independent of $p$ and $T$, and provide an order of magnitude
(0.6 down to 0.06) improvement in $\nu$ over the pure quantum value.
Decoherence just on the coin does not enhance the uniformity of the
distribution, as Fig. \ref{linedist} shows, there is a cusp at
$x=0$.

\subsection{Decoherence in a Quantum Walk on a $N$-Cycle}\label{decycle}

\begin{figure}[t!]
    \begin{center}
	\includegraphics[width=0.7\columnwidth]{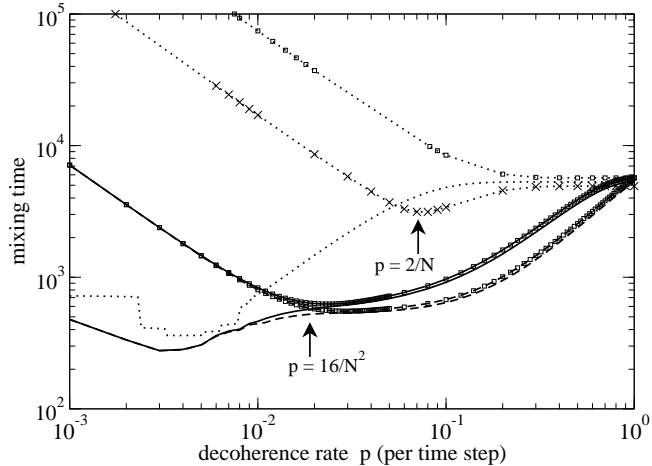}
    \end{center}
    \caption{Numerical data for mixing times on cycles of size
	$N=29$ and $N=30$ ($\square$), for coin (dotted), particle (dashed)
	and both (solid) subject to decoherence, using
	$\varepsilon=0.01$. Also $N=28$ ($\times$) for coin.
	Both axes logarithmic} 
    \label{mixcycle}
\end{figure}

We now consider a walk on a $N$-cycle subjected to decoherence.
There is an experimental proposal for implementation of a quantum walk
on a cycle in the phase of a cavity field \cite{sanders02a}, in which
further aspects of decoherence in such quantum walks are considered.
Recall from Sect. \ref{ex_lorc} that the
pure quantum walk on a cycle with $N$ odd, is known \cite{aharonov00a}
to mix in time $\le O(N\log N)$ if a Hadamard coin is used. The
quantum walk on a cycle with $N$ 
even does not mix to the uniform distribution with a Hadamard coin,
but can be made to do so by appropriate choice of coin flip operator
\cite{tregenna03a}.
Under the action of a small amount of decoherence, the mixing time becomes
shorter for all cases, typical results are shown in Fig.~\ref{mixcycle}.
In particular, decoherence causes the even-$N$ cycle to mix to the
uniform distribution even when a Hadamard coin is used.
The asymptotes in Fig.~\ref{mixcycle} for $N$ even and decoherence on
the coin only, for $p < 2/N$, are well fitted by
$\varepsilon p M_{\varepsilon} \simeq N/4$ for $N$ divisible by 2, and
$\varepsilon p M_{\varepsilon} \simeq N/16$ for $N$ divisible by 4.
For larger $p$, the mixing time tends to the classical value
of $N^2/16\varepsilon$.
Although for $N$ divisible by 4, the (coin-decohered)
mixing time shows a minimum below the
classical value at $p \simeq 2/N$, this mixing time is still quadratic in
$N$.  Thus although noise on the coin causes the even-$N$ cycle to mix
to the uniform distribution, it does not produce a significant speed up over
the classical random walk.

For decoherence on the particle position, with $p < 16/N^2$,
$\varepsilon p M_{\varepsilon} \simeq 1/(N/2-1)$ for $N$ divisible by 2, and
$\varepsilon p M_{\varepsilon} \simeq 1/(N/4+3)$ for $N$ divisible by 4.
At $p \simeq 16/N^2$, there is a minimum in the mixing time at a value
roughly equal to the $(N+1)$-cycle pure quantum mixing time,
$M^{(\text{min})}_{\varepsilon} \sim \alpha N/\varepsilon$ (with $\alpha$ a
constant of order unity).
Decoherence on the particle position thus causes the even-$N$ cycle to
mix in linear time for a suitable choice of decoherence rate $p^{(\text{min})}
\sim 16/N^2$, independent of $\varepsilon$ so long as $\varepsilon < 1/N$.

For all types of decoherence, the odd-$N$ cycle shows a minimum mixing
time at a position somewhat earlier than the even-$N$ cycle,
roughly $p = 2/N^2$,
but because of the oscillatory nature of $\overline{P(x,p,T)}$,
the exact behaviour \cite{kendon02b} is not a smooth function
of $p$ or $\varepsilon$.  
As decoherence on the particle (or both) increases, at $p \simeq 16/N^2$,
the mixing time passes through an inflexion and from then on behaves
in a quantitatively similar manner to the adjacent-sized even-$N$ cycles,
including scaling as $M^{(\text{min})}_{\varepsilon} \sim \alpha N/\varepsilon$
at the inflexion.  Thus for at least $0 \le p \lesssim 16/N^2$ 
the mixing time stays linear in $N$, and exhibits the quantum speed up over
the classical $N^2$.

\subsection{Decoherence in a Quantum Walk on a Hypercube}

Recalling from Sect. \ref{ex_hype}, we are interested in the
hitting time to the opposite corner, which was shown by
Kempe \cite{kempe02a} to be polynomial, an exponential speed up
over a classical random walk.
Kempe discusses two types of hitting times, one-shot,
where a measurement is made after a pre-determined number of steps,
and concurrent, where the desired location is monitored
continuously to see if the particle has arrived.
In each case, the key parameter is the probability $P_h$ of
finding the particle at the chosen location. Here we consider only
target locations exactly opposite the starting vertex.
We have calculated $P_h$ numerically following the scheme of
\cite{moore01a,kempe02a} with a $N$-dimensional coin
and the Grover coin operator, defined in (\ref{grovercoin}). 
\begin{figure}[t!]
    \begin{center}
	\includegraphics[width=0.7\columnwidth]{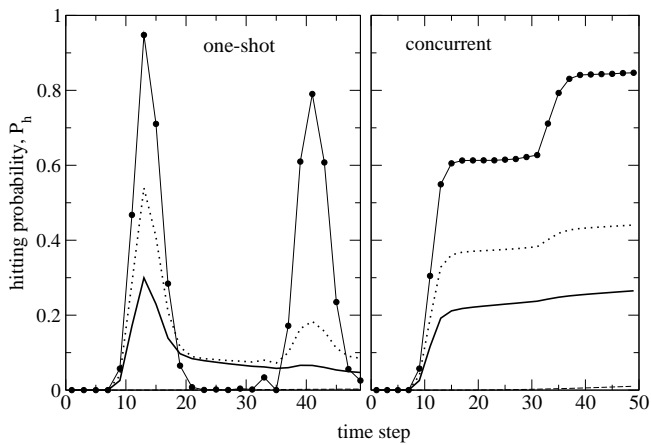}
    \end{center}
    \caption{Hitting probability on a 9--dimensional hypercube for
	one-shot (left) and concurrent (right), perfect walk (circles),with $p
	= 0.05$ (dotted), $p = 0.1 \simeq 1/9$ (solid). Classical hitting
	probability barely visible (dashed)}
    \label{dechit}
\end{figure}
Figure \ref{dechit} shows how $P_h$ is affected by decoherence.
All forms of decoherence have a similar effect on $P_h$, reducing the
peaks and smoothing out the troughs.
For the one-shot hitting time this is useful, 
raising $P_h$ in the trough to well above the classical value,
so it is no longer necessary to know exactly when to measure.
For $p \lesssim 1/N$, the height of the first peak scales as
$P_h(p) = P_h(0) \exp\{-(N+\alpha)p\}$,
where $0\lesssim\alpha\lesssim 2$ depending on whether coin, particle
or both are subject to decoherence.
Thus $P_h$ decreases exponentially in $p$, but
$p\simeq 1/N$ only lowers $P_h$ by a factor of $1/e$,
still exponentially better than classical.
Continuous monitoring of the target location as in the concurrent
hitting time is already a type of controlled decoherence,
no new features are produced by the addition of unselective decoherence,
but there is still a range of $0 < p \lesssim 1/N$ within which
the quantum speed up is preserved.

\subsection{Summary of Decoherence Effects in Coined Quantum Walks}

For the walk on a line, whilst the effect of decoherence is to reduce
the standard deviation rapidly towards the classical value,
it is possible to generate highly uniform distributions
over a range proportional to $T$ for small values of the decoherence
parameter, $p$.  Uniform sampling is one of the basic tasks for which 
classical random walks are used, so being able to do this over a
quadratically larger range with a quantum walk is certainly
promising, though no quantum algorithms using this have been
described to date.
For a walk on a cycle the effects of decoherence are
more beneficial.  An optimum rate of decoherence exists for which the
rate of mixing is enhanced beyond the pure quantum bound.
Further, any amount of decoherence removes the effect of the
coin flip operator on the steady state,
allowing all such walks to converge to a uniform distribution.
Fast mixing to a uniform distribution is again an important
basic property required for efficient random sampling, and is the limiting
factor in many classical algorithms.
For a hypercube, decoherence can still be tolerated so
long its rate is kept smaller than $O(1/N)$, which is logarithmic in the
system size ($2^N$).

Thus, for both experiments and algorithms, decoherence need only
be controlled down to finite low levels, rather than negligible levels,
in order to observe the intriguing quantum effects
displayed by coined quantum walks.
Many open problems remain concerning the best ways to do this, 
how to exploit the full power of the extra degrees of freedom
provided by the quantum coin, and how to make quantum walks perform
useful tasks over more general graphs that provide real quantum
computational advantages over classical algorithms.


\end{document}